\documentclass[%
 reprint,
 amsmath,amssymb,
 aps,
]{revtex4-2}
\usepackage[colorlinks=true, allcolors=blue]{hyperref}
\usepackage{graphicx}% Include figure files
\usepackage{dcolumn}% Align table columns on decimal point
\usepackage{bm}% bold math
\usepackage{here}

\newcommand{\jcap}{Journal of Cosmology and Astroparticle Physics}
\newcommand{\highlight}[1]{#1}
\begin{document}

\preprint{APS/123-QED}

%\title{Manuscript Title:\\with Forced Linebreak}% Force line breaks with \\

\title{Collisional and Fast Neutrino Flavor Instabilities in Two-dimensional Core-collapse Supernova Simulation with Boltzmann Neutrino Transport}

\author{Ryuichiro Akaho}
\affiliation{Graduate School of Advanced Science and Engineering, Waseda University, 3-4-1 Okubo, Shinjuku, Tokyo 169-8555, Japan}

\author{Jiabao Liu}
\affiliation{
Graduate School of Advanced Science and Engineering, Waseda University, 3-4-1 Okubo, Shinjuku, Tokyo 169-8555, Japan}

\author{Hiroki Nagakura}
\affiliation{Division of Science, National Astronomical Observatory of Japan, 2-21-1 Osawa, Mitaka, Tokyo 181-8588, Japan}

\author{Masamichi Zaizen}
\affiliation{Faculty of Science and Engineering, Waseda University, 3-4-1 Okubo, Shinjuku, Tokyo 169-8555, Japan}

\author{Shoichi Yamada}
\affiliation{Advanced Research Institute for Science and Engineering, Waseda University, 3-4-1 Okubo, Shinjuku, Tokyo 169-8555, Japan}

\date{\today}% It is always \today, today,

\begin{abstract}
We present a comprehensive study on the occurrences of 
the collisional flavor instability (CFI) and
the fast flavor instability (FFI) of neutrinos based on a two-dimensional (2D) core-collapse supernova (CCSN) simulation performed with a Boltzmann radiation hydrodynamics code. 
We find that CFI occurs in a region with the baryon-mass density of
$10^{10}\lesssim \rho \lesssim 10^{12}\,\mathrm{g}\,\mathrm{cm}^{-3}$, which is similar to the previous results in one-dimensional (1D) CCSN models.
In contrast to 1D, however, the CFI region varies with time vigorously in the 2D model, whereas it had a quiescent structure in 1D. This is attributed to the fact that the turbulent flows advected from a gain region account for the temporal variations. 
Another noticeable difference from the 1D models is the appearance of resonance-like CFI where number densities of $\nu_e$, $\bar\nu_e$ nearly coincide each other. The CFI growth rate there is enhanced and can reach \highlight{$\sim10^{8}\,\mathrm{s}^{-1}$}.
As for FFI, on the other hand, it appears in three different regions;
(1) the region overlapped with the resonance-like CFI,
(2) neutrino decoupling regions where $\bar{\nu}_e$'s are strongly emitted, and 
(3) optically thin regions where neutral-current scatterings dominate over charged-current reactions.
Although overall properties for FFI are consistent with previous studies, 
we find that the number of electron-neutrinos lepton number crossing (ELN crossing) temporary becomes multiple, which can be assessed accurately only by multi-angle treatments in neutrino transport.
We find that the growth rate of FFI is always higher than CFI if both of them occur, which suggests that the former is dominant for the linear evolution.

%\begin{description}
%\end{description}
\end{abstract}

\maketitle

%\tableofcontents

\section{Introduction}
Core-collapse supernovae (CCSNe) are known to occur at the end of massive star's lives. Although the quest for physical process of explosion mechanism is still ongoing, detailed numerical simulations have suggested
that neutrinos play key roles in driving explosions (see a recent review \cite{Burrows2021}). One thing we need to mention is that most CCSN simulations assumed that neutrino oscillations (or flavor conversions) do not occur in post-shock regions due to a high-density matter suppression. However, 
neutrinos are also dense in the environment, that can offer another channel to induce neutrino flavor conversions (see \cite{Duan2010,Tamborra2021} for reviews). Similar as the Mikheyev Smirnov Wolfenstein mechanism, neutrino self-interactions can also induce refractive effects. If a certain condition is met, flavor instabilities can take place. 
In this study, we will present evidence that two distinct flavor instabilities can occur ubiquitously in post-shock regions of CCSN core, based on a recent two-dimensional (2D) CCSN model.

One of the collective neutrino oscillation models, referred to as the fast flavor instability (FFI) \cite{Sawyer2005,Sawyer2009,Sawyer2016}, has been attracting great attention.
It is expected to evolve in a very short timescale, and many studies have been performed extensively to investigate its outcome \cite{Pantaleone1992,Chakraborty2016,Abbar2018,Dasgupta2018a,Capozzi2019,Johns2020,Martin2020,Shalgar2021,Padilla-Gay2021b,Richers2021a,Richers2021b,Tamborra2021,Wu2021,Bhattacharyya2022,Kato2022,Richers2022a,Richers2022b,Sasaki2022,Fiorillo2023a,Fiorillo2023b,Xiong2023b,Zaizen2023}.
Many post-processing analyses of CCSN simulations demonstrated that FFI is likely to appear in CCSNe \cite{Abbar2019,DelfanAzari2019,Nagakura2019b,Abbar2020a,Nagakura2021b,Harada2022,Akaho2023}. The FFI is also studied in the context of binary neutron star merger remnant \cite{George2020,Li2021,Padilla-Gay2021a,Just2022,Fernandez2022,Richers2022c,Grohs2023a,Grohs2023b,Nagakura2023c}.
The FFI is known to take place when the neutrino flavor lepton number (NFLN) crosses zero \cite{Morinaga2022,Dasgupta2022}. Mathematically we need full information on the neutrino distributions in momentum space, such as the one provided by the simulation with the Boltzmann neutrino transport.
It is worth mentioning that several methods have been proposed to infer the existence of angular crossing only from truncated moments \cite{Dasgupta2018b,Glas2020,Abbar2020b,Nagakura2021a,Nagakura2021b,Johns2021,Abbar2023}.

Effects of FFI on the CCSN dynamics is not fully understood yet. 
In \cite{Nagakura2023a}, the space-dependent and fully nonlinear flavor conversion was calculated by directly solving the quantum kinetic equation (QKE) for a realistic CCSN background. It was demonstrated that the flavor conversion could greatly reduce the neutrino heating in the gain region because electron-type neutrinos are mixed with the heavy-leptonic neutrinos, which has lower number fluxes.

Possible feedback of the flavor conversions to CCSN dynamics were studied both in 1D \cite{Ehring2023a} and 2D \cite{Ehring2023b}. In these papers, the criterion for the flavor mixing was parametrized by the matter density instead of the neutrino distribution itself. They observed that the flavor conversions in the gain region enhance matter heating considerably whereas those occurring in the proto-neutron star (PNS) accelerates the cooling. 
Although their numerical method needs to be improved for more physically accurate models of FFI,
these results clearly indicate that the FFI could have a large impact on the CCSN dynamics.

Recently, a new type of instability called collisional flavor instability (CFI) \cite{Johns2023} was discovered. It is driven by collisions, which are considered on the right hand side of the Boltzmann equation, and does not require the NFLN crossing as FFI does. 
As such, the CFI growth rate is determined by the collision rates except when the resonance-like phenomenon \cite{Lin2023,Liu2023a,Xiong2023c} occurs, in which case the growth rate is enhanced, being proportional to the square-root of the neutrino number density.
The nonlinear dynamics of CFI has been recently studied and many interesting features have been discovered \cite{Padilla-Gay2022,Johns2022,Lin2023,Xiong2023a,Xiong2023c,Kato2023b}.
For example, it was recently found that the resonance-like CFI may lead to the flavor swap \cite{Kato2023b}.
Moreover, the interplay of CFI and FFI may enhance the flavor conversion \cite{Johns2022,Padilla-Gay2022} (but also see \cite{Hansen2022}).
The possibility of CFI in realistic CCSN models was first investigated in \cite{Xiong2023a}. Very recently, a systematic study of CFI for various progenitors was conducted based on 1D CCSN models \cite{Liu2023b}. The results of these studies are both affirmative.

The study on the possible occurrence of CFI in realistic CCSN simulations are so far limited to 1D \cite{Xiong2023a,Liu2023a}.
The conclusions may change qualitatively in multi-dimensions. 
In addition, the purpose of this study is to assess the occurrence of CFI and FFI simultaneously. Multi-dimensional simulation is crucial because the FFI
tends to be suppressed in 1D \cite{Tamborra2017}
for the following reason. In 1D, the deleptonization is slower due to the absence of convection, and $Y_e$ tends to become higher, which makes $\nu_e$ dominant over $\bar{\nu}_e$ in the entire region.

In this paper, we perform the post-process analyses of CFI and FFI simultaneously for our 
2D CCSN model performed with the Boltzmann radiation hydrodynamics code. We judge the occurrence of these flavor conversions and estimate their linear growth rates based on the analytical formulae we derived in the previous studies, \cite{Liu2023a} for CFI and \cite{Morinaga2020} for FFI. In the search of FFI, we employ the full information on the angular distributions of neutrinos in momentum space, that is obtained in the Boltzmann transport. This is a great advantage over other analyses based on approximate neutrino transport.

This paper is organized as follows.
Section \ref{sec_linear} explains how we evaluate the linear growth rates of CFI and FFI. We also give a short explanation of the CCSN simulation model analyzed in this study.
We present the results of the analyses of CFI and FFI in 2D CCSN simulation in section \ref{sec_results}, and the conclusions are given in section \ref{sec_conclusion}.
we use the metric signature of $+ - - -$. The natural unit $c = \hbar = 1$ is employed, where $c$ and $\hbar$ denote the speed of light, and the Planck constant, respectively. 

\section{Flavor Instabilities and CCSN Model}
\label{sec_linear}
\subsection{Collisional Flavor Instability}
In this paper, we employ the analytical formulae we derived in our previous linear analysis on CFI \cite{Liu2023a,Liu2023b}. In the following, we give a quick look at the derivation. For more details we refer readers to the original papers \cite{Liu2023a,Liu2023b}.
We start from the QKE
\begin{equation}
iv^\mu\cdot\partial_\mu\rho=[H,\rho]+iC,
\end{equation}
where $v^\mu$, $\rho$, $H$ and $C$ denote the neutrino four-velocity, density matrix, Hamiltonian, and the collision terms, respectively. 
We work in the two flavor framework, where the $\mu$-type neutrino ($\nu_\mu$) and the $\tau$-type neutrino ($\nu_\tau$) are assumed to be identical and denoted as $\nu_x$.
The components of the density matrix are represented as
\begin{equation}
\rho\equiv
\left(
\begin{matrix}
f_{\nu_e} & S_{ex} \\
S_{xe} & f_{\nu_x} 
\end{matrix}\right).
\end{equation}
\highlight{They are functions of the spatial position $x$ and the four-momentum $P$}. 
The Hamiltonian, which has the contributions from the vacuum, matter and neutrino self-interactions, is given as follows: 
\begin{eqnarray}
H\equiv\frac{M^2}{2E}&+&\sqrt{2}G_F v_\mu\mathrm{diag}(j_e^\mu(x),j_x^\mu(x))\nonumber\\&+&\sqrt{2}G_Fv_\mu\int dP'\rho(x,P')v'^\mu,
\end{eqnarray}
where $M^2$, $j^\mu_\alpha(x)$ and $G_F$ are the neutrino mass-squared matrix, the lepton number four-currents, and the Fermi constant, respectively.
The integral over momentum is expressed as 
\begin{equation}
\int dP\equiv\int_{-\infty}^{\infty}\frac{E^2dE}{2\pi^2}\int\frac{d\Omega_p}{4\pi},
\end{equation}
where $E$ and $\Omega_p$ are the energy and solid angle in momentum space, respectively. \highlight{Following the common practice, the negative energy corresponds to antineutrinos.}
Throughout this paper, we ignore the vacuum term.
The collision term is given in the relaxation approximation \cite{Johns2023} as 
\begin{equation}
C(x,P)\equiv\frac{1}{2}\{\mathrm{diag}(\Gamma_{\nu_e}(x,P),\Gamma_{\nu_x}(x,P)),\rho_\mathrm{eq}-\rho\}.
\end{equation}
where $\Gamma_{\nu_\alpha}(x,P)$ and $\rho_\text{eq}$ stand for the collision rates and the density matrix for the equilibrium state, respectively. The curly bracket denotes the anti-commutator. 
We take into account all emission/absorption interactions incorporated in the CCSN simulation model (see section \ref{sec_model}), but scattering reactions are neglected.
In our analyses, we treat angular-integrated distribution for CFI and the scatterings are considered to be exactly cancelled (see also \cite{Liu2023b} for more details). \highlight{Note that our assumption may not be correct in the optically-thin regions where the neutrino distribution greatly deviates from isotropic. However, the timescale of neutrino-matter interactions also becomes longer there, indicating that CFI would not grow appreciably. Hence, this region is not of interest in CFI.}

Assuming $S_{ex}\ll f$, we linearlize the QKE for the off-diagonal component. Adopting the plane wave ansatz:
$S_{ex}(x,P)=\tilde{S}_{ex}(x,k)e^{ikx}$,
we obtain the following homogeneous equation
\begin{equation}
\label{eq_pi_a}
\Pi_{ex}^{\mu\nu}(k)a_\nu(k)=0,
\end{equation}
where the matrix $\Pi_{ex}^{\mu\nu}(k)$ and the vector $a^\mu(k)$ are defined as
\begin{equation}
\Pi_{ex}^{\mu\nu}(k)=\eta^{\mu\nu}+\sqrt{2}G_F\int dP\frac{(f_{\nu_e}-f_{\nu_x})v^\mu v^\nu}{v^\lambda(k_\lambda-\Lambda_{0e\lambda}+\Lambda_{0x\lambda})+i\Gamma_{ex}},
\end{equation}
\begin{equation}
a^\mu(k)\equiv\sqrt{2}G_F\int\mathrm{d}P\tilde{S}_{ex}(k,P)v^\mu,
\label{eq_vec_a}
\end{equation}
where $\Lambda_{0\alpha}$ is defined as
\begin{equation}
\Lambda^\mu_{0\alpha}\equiv\sqrt{2}G_F[j^\mu_{\alpha}(x)+\int\mathrm{d}Pf_{\nu_\alpha}(x,P)v^\mu].
\end{equation}
Since $\Lambda$'s are real, they do not affect the instability and can be absorbed into $k$ in the denominator of $\Pi_{ex}$.
The nontrivial solution exists if and only if the following relation is satisfied:
\begin{equation}
\det\Pi_{ex}(k)=0.
\end{equation}
Note that Eq. \ref{eq_pi_a} is not limited to CFI but FFI can be treated equally.

In order to consider CFI alone, we assume that the neutrino distribution is isotropic. NFLN crossing, the condition for FFI, is assumed to be absent from the beginning. In addition, we only consider the case $k=0$. Then only the tensor $v\otimes v$, which is a collection of trigonometric functions (see Eq. 14 in \cite{Liu2023a}), depends on the angle $\Omega_p$ in momentum space. Only the diagonal components survive after integration, and the three spatial components are degenerate. The resultant equation is
\begin{equation}
\label{eq_I}
I\equiv\sqrt{2}G_F\int_{-\infty}^{\infty}\frac{E^2\mathrm{d}E}{2\pi^2}\frac{f_{\nu_e}(E)-f_{\nu_x}(E)}{\omega+\mathrm{i}\Gamma_{ex}(E)}=-1,3.
\end{equation}
The case for $I=-1$ corresponds to the time ($0$-th) component and that for $I=3$ comes from the spatial components.
Further assuming that the neutrino distribution is monochromatic
\begin{equation}
f_{\nu_e}(E)-f_{\nu_x}(E)=\frac{2\pi^2}{\sqrt{2}G_F E^2}[{\mathfrak{g}}\delta(E-\epsilon)-\bar{{\mathfrak{g}}}\delta(E+\bar{\epsilon})],
\end{equation}
we obtain the following reduced equations
\begin{equation}
\label{eq_I_2}
\frac{\mathfrak{g}}{\omega+i\Gamma}-\frac{\bar{\mathfrak{g}}}{\omega+i\bar{\Gamma}}=-1,3,
\end{equation}
where $\mathfrak{g}$, $\bar{\mathfrak{g}}$ are defined as
\begin{equation}
\mathfrak{g}\equiv n_{\nu_e}-n_{\nu_x},\;
\bar{\mathfrak{g}}\equiv n_{\bar{\nu}_e}-n_{\bar{\nu}_x},\;
\end{equation}

The solution for $I=-1$, called the isotropy-preserving branch, is given as 
\begin{equation}
\label{eq_omega_p}
\omega_\pm^\mathrm{pres}=-A-i\gamma\pm\sqrt{A^2-\alpha^2+2iG\alpha},
\end{equation}
and the solution for $I=3$, called the isotropy-breaking branch, is given as
\begin{equation}
\label{eq_omega_b}
\omega_\pm^\mathrm{break}=-\frac{A}{3}-i\gamma\pm\sqrt{(\frac{A}{3})^2-\alpha^2-\frac{2}{3}iG\alpha}.
\end{equation}
Symbols $G$, $A$, $\gamma$, $\alpha$ are defined as
\begin{equation}
\label{eq_G_A}
G\equiv\frac{\mathfrak{g}+\bar{\mathfrak{g}}}{2},\; 
A\equiv\frac{\mathfrak{g}-\bar{\mathfrak{g}}}{2},\;
\gamma\equiv\frac{\Gamma+\bar{\Gamma}}{2},\;
\alpha\equiv\frac{\Gamma-\bar{\Gamma}}{2},
\end{equation}
where the collision rates $\Gamma$, $\bar\Gamma$ are given as
\begin{equation}
\label{eq_gamma}
\Gamma\equiv\frac{\Gamma_e+\Gamma_x}{2},\;
\bar\Gamma\equiv\frac{\bar{\Gamma}_e+\bar{\Gamma}_x}{2},
\end{equation}
and $n_{\nu_i}$ and $\Gamma_i$ are the number densities, and the energy-integrated collision rates, respectively. They are expressed as follows:
\begin{equation}
\label{eq_numberdensity}
n_i=\sqrt{2}G_F\int \frac{E^2dE}{2\pi^2} f(E),
\end{equation}
\begin{equation}
\label{eq_gamma_i}
\Gamma_i\equiv\sqrt{2}G_F\int \frac{E^2 dE}{2\pi^2} \Gamma(E)f_i(E),
\end{equation}
with $\Gamma(E)$ being the energy-dependent emission/absorption rates. 

CFI occurs when the imaginary part of $\omega$ is positive. Eqs. \ref{eq_omega_p} and \ref{eq_omega_b} are obtained under the assumption that the neutrino distribution is isotropic and monochromatic. In \cite{Liu2023a}, it was found that they are reasonable approximations if the average energies of neutrino and anti-neutrino are plugged in $\epsilon$ and $\bar{\epsilon}$, respectively, as long as there is no NFLN crossing.

In this paper, we define the CFI growth rate as
\begin{equation}
\label{eq_growth}
\sigma_\mathrm{CFI}\equiv\mathrm{max}\left(\mathrm{Im}(\omega_\pm^\mathrm{pres}),\mathrm{Im}(\omega_\pm^\mathrm{break})\right).
\end{equation}
\highlight{The growth rate can be calculated by using the number densities (Eq. \ref{eq_numberdensity}) and the energy-integrated collision rates Eq. \ref{eq_gamma_i}), which are provided by the CCSN model.}

It is useful to consider the following limits:
\highlight{
\begin{equation}
\text{max\,(Im\,}\omega_\pm^\mathrm{pres})=\begin{cases}
    -\gamma+\frac{|G\alpha|}{|A|},& \quad (A^2\gg |G\alpha|),\\
    -\gamma+\sqrt{|G\alpha|},& \quad (A^2\ll |G\alpha|),
\end{cases}
\end{equation}
}
for the isotropy-preserving branch and
\highlight{
\begin{equation}
\text{max\,(Im\,}\omega_\pm^\mathrm{break})=\begin{cases}
    -\gamma+\frac{|G\alpha|}{|A|},&\quad (A^2\gg |G\alpha|),\\
    -\gamma+\frac{\sqrt{|G\alpha|}}{\sqrt{3}},& \quad (A^2\ll |G\alpha|),
\end{cases}
\end{equation}
}
for the isotropy-breaking branch.

In the typical CCSN situation, $A^2 \gg |G\alpha|$ is satisfied because $A\sim G\gg\alpha$ \cite{Liu2023b}. However, if $n_{\nu_e}$ and $n_{\bar{\nu}_e}$ are very close to each other, $A$ becomes small and the lower case ($A^2\ll|G\alpha|$) applies. Then the growth rate is $\sim\sqrt{G\alpha}$, which is larger than the ordinary CFI growth rate of $\sim G|\alpha|/A$. This is called the resonance-like CFI \cite{Lin2023,Liu2023a,Xiong2023c}. In contrast, we will refer to the CFI in the regime of $A^2\gg |G\alpha|$ as the ``non-resonance" CFI hereafter.

\subsection{Fast Flavor Instability}
As mentioned previously, existence of FFI is known to occur when NFLN crossing is present in the neutrino angular distribution in momentum space \cite{Morinaga2022,Dasgupta2022}. Since we employ the 3-species neutrino transport for $\nu_e$, $\bar{\nu}_e$ and $\nu_x$ in the CCSN simulation (see section \ref{sec_model} below), we have only to look at the electron lepton number (ELN) crossing i.e., the crossing between $\nu_e$ and $\bar\nu_e$.
In this paper, the FFI growth rate is estimated based on the following empirical formula proposed in \cite{Morinaga2020}:
\begin{equation}
\sigma_\mathrm{FFI}\equiv\sqrt{-\left(\int_{\Delta G>0}\frac{d\Omega_p}{4\pi}\Delta G\right)\left(\int_{\Delta G<0}\frac{d\Omega_p}{4\pi}\Delta G\right)},
\end{equation}
where $\Delta G$ is defined as 
\begin{equation}
\Delta G \equiv \sqrt{2}G_F\int\frac{E^2dE}{2\pi^2}\left(f_{\nu_e}-f_{\bar{\nu}_e}\right).
\end{equation}
It is apparent that the above formula gives nonzero positive values if and only if there is at least one angular crossing. The formula was also used in our previous studies of FFI in Boltzmann CCSN simulations \cite{Nagakura2019b,Harada2022,Akaho2023}.

\subsection{CCSN Model}
\label{sec_model}
We give here only basic information on the CCSN model we employ in this paper.
It is a result of the 2D
CCSN simulation under axisymmetry for the progenitor with the zero-age main sequence mass of $11.2M_\odot$ \cite{Woosley2002}. 
The Boltzmann equations are faithfully solved for three neutrino species ($\nu_e$, $\bar{\nu}_e$ and $\nu_x$) by discretizing the entire phase space, i.e., by the $S_N$ method.
Newtonian hydrodynamics equations are solved simultaneously with the feedback from/to neutrinos fully taken into account. 
The radial range of $[0:5000]\,\mathrm{km}$ is divided into 384 grid points, and the zenith angle $\theta\in[0:\pi]$ is divided into 128 grid points. The energy range of $[0:300]\,\mathrm{MeV}$ is divided into 20 logarithmically spaced grid points. The zenith angle in momentum space $\theta_\nu\in[0:\pi]$ and the azimuth angle $\phi_\nu\in[0:2\pi]$ are divided into 10, and 6 grid points, respectively.
The neutrino-matter interactions are based on the so-called standard set
\cite{Bruenn1985} with a few modifications; the inelastic scattering off electrons and the nucleon–nucleon bremsstrahlung \cite{Friman1979} are implemented. Note that the emission/absorption rates in Eq. \ref{eq_gamma_i} for the CFI growth rates are the same as those employed in the simulation.
The details of the numerical code are described in our series of papers \cite{Nagakura2014,Nagakura2017,Nagakura2019a}. In this simulation, Lattimer-Swesty equation of state \cite{Lattimer1991} with the incompressibility parameter $K=220\mathrm{MeV}$ is employed.
The simulation was conducted up to $\sim400\,\mathrm{ms}$ after bounce when we observed a successful explosion with the maximum shock radius reaching $1000\,\mathrm{km}$ in $t\sim400\,\mathrm{ms}$ after bounce. See \cite{Harada2020} for the details of this simulation. 

\section{Results}
\label{sec_results}
\subsection{Overall Properties}
\begin{figure*}[t] 
\includegraphics[width=\linewidth]{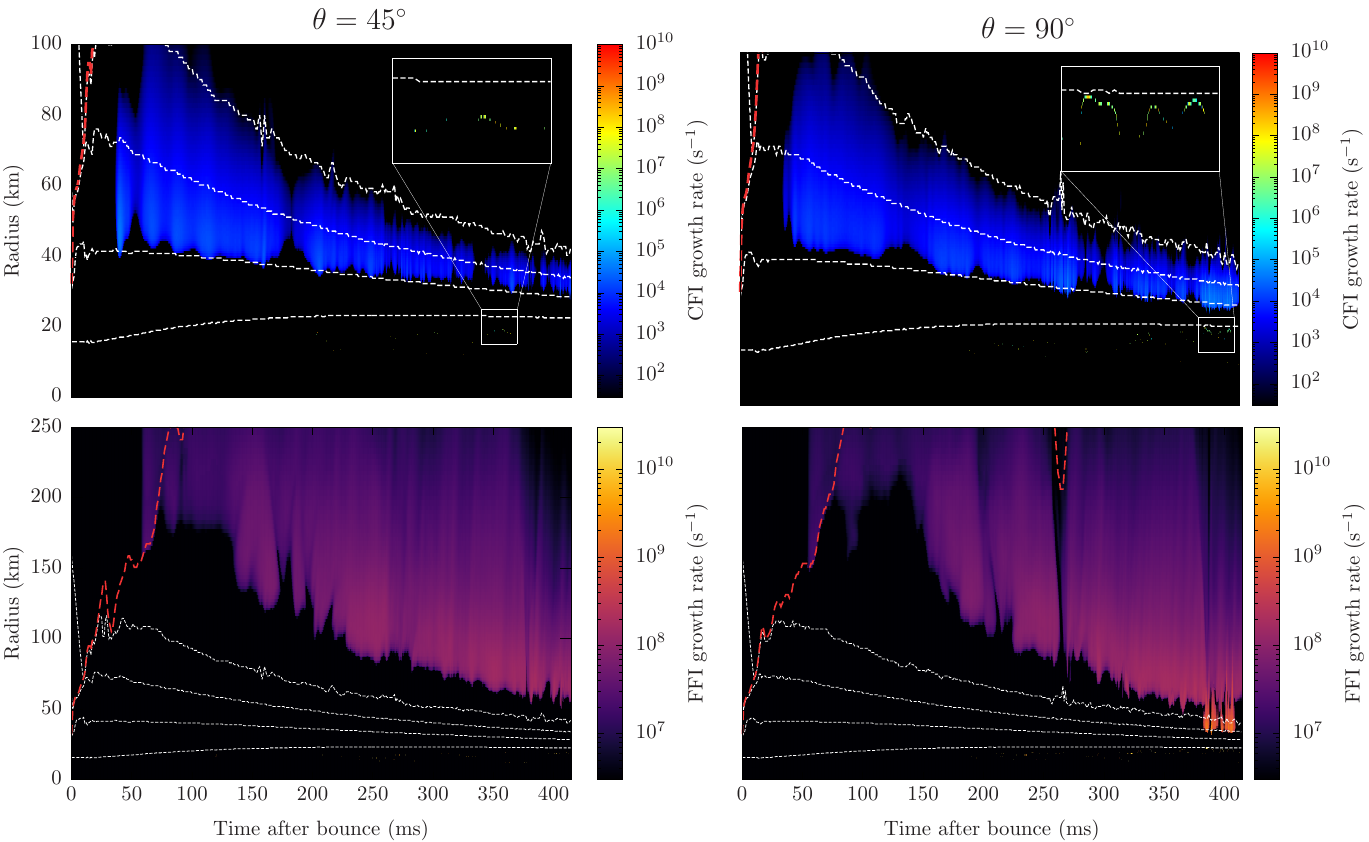}
\caption{\highlight{Time-radius map of the growth rate of CFI (top) and FFI (bottom) for the angle $\theta=45^\circ$ and $90^\circ$}. White broken lines, from top to bottom, denote the radius for the density $10^{10}$, $10^{11}$, $10^{12}$ and $10^{13}\mathrm{g}\,\mathrm{cm}^{-3}$, respectively. Red broken line denote the shock radius.}
\label{fig_TRmap_CFI}
\end{figure*}
\begin{figure*}[t]
\includegraphics[width=\linewidth]{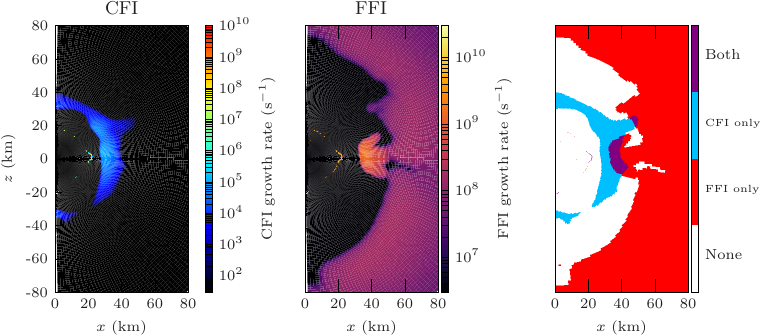}
\caption{Meridian map of CFI growth rate (left), FFI growth rate (middle), and the dominant instability (right) at $t=404\mathrm{ms}$ after bounce.}
\label{fig_2Dslices}
\end{figure*}

Top panels of Fig. \ref{fig_TRmap_CFI} shows the time-radius maps of CFI growth rate at $\theta=45^\circ$ and $90^\circ$. 
CFI is expected to occur in the region with a bright color. In fact, the black regions in the plots have growth rates smaller than $10^{-9}\,\mathrm{cm}^{-1}$, and we do not think CFI is important there. It is clear at both angles (and actually at all angles as shown in Fig. \ref{fig_2Dslices}) that a CFI region appears at $t\sim50\,\mathrm{ms}$ for the first time and continues to exist later on. This unstable region moves to smaller radii as the PNS contracts. 
It roughly corresponds to the region with $10^{10}\lesssim \rho \lesssim 10^{12}\,\mathrm{g\,}\mathrm{cm}^{-3}$, similar density range as reported in a 1D study \cite{Liu2023b}. In the 2D case, however, the radial extent of the region changes rather rapidly in time whereas such time variations were absent in the 1D model. 
This is due to the turbulence that occurs commonly on the multi-dimensional models.

A closer inspection of the  plots reveals another CFI region deeper inside, $r \sim 20\,\mathrm{km}$, at later times, $t \gtrsim 200\,\mathrm{ms}$, (see the magnified figures). It is very narrow but has greater growth rates than the region mentioned above and was not found in the 1D model. In fact, this corresponds to the resonance-like CFI, a feature unique to multi-dimensional models, as we discuss later. 

For comparison we present the time-radius maps of the FFI growth rate in the bottom panels of Fig. \ref{fig_TRmap_CFI}. The reddish region is unstable to FFI this time. Note that the radial range and the color scale are different between top and bottom panels. The four dashed lines that show the locations of $\rho=10^{10}$, $10^{11}$, $10^{12}$, $10^{13}\,\mathrm{g}\,\mathrm{cm}^{-3}$ will help the correspondence between the plots.
There is a wide FFI region with located at larger radii much outside than the CFI region in general. In the late phase, $t\gtrsim400\,\mathrm{ms}$ however, the two regions are partially overlapped with each other at $\theta=90^{\circ}$. Note that we analyze CFI and FFI independently, assuming that the latter is absent in the analysis of the former.

The spatial extents of the CFI and FFI regions in the meridian section are shown in Fig. \ref{fig_2Dslices} at $t=404\,\mathrm{ms}$ after bounce. 
The resonance-like CFI occurs sporadically at $r\sim20\,\mathrm{km}$ whereas the non-resonance CFI regions prevail at $30\lesssim r\lesssim40\,\mathrm{km}$.
The FFI region is extended at even larger radii, $r \gtrsim 50\,\mathrm{km}$, but also appears at almost the same positions as the resonance-like CFI. Although the non-resonance CFI region is mostly separated from the FFI region, there are some overlaps (see the rightmost panel of Fig. \ref{fig_2Dslices}). It is apparent that it occurs in a convective eddy. 
\highlight{The growth rates of CFI and FFI tend to be higher around the equator than near the poles. This comes from the stronger $\bar{\nu}_e$ emission in the lower latitudes, induced by the large-scale fluid motion.
The morphology of fluid motion is known to be qualitatively different between 2D and 3D \cite{Couch2013}, and the degree of asymmetry may be exaggerated in this study. However, the qualitative trend will be unchanged in 3D.}

In the following subsections \ref{sec_CFI} and \ref{sec_FFI}, we look into CFI and FFI individually.
The growth rates of CFI and FFI are compared in subsection \ref{sec_comparison}.

\subsection{CFI}
\label{sec_CFI}
\begin{figure*}[t] 
\includegraphics[width=\linewidth]{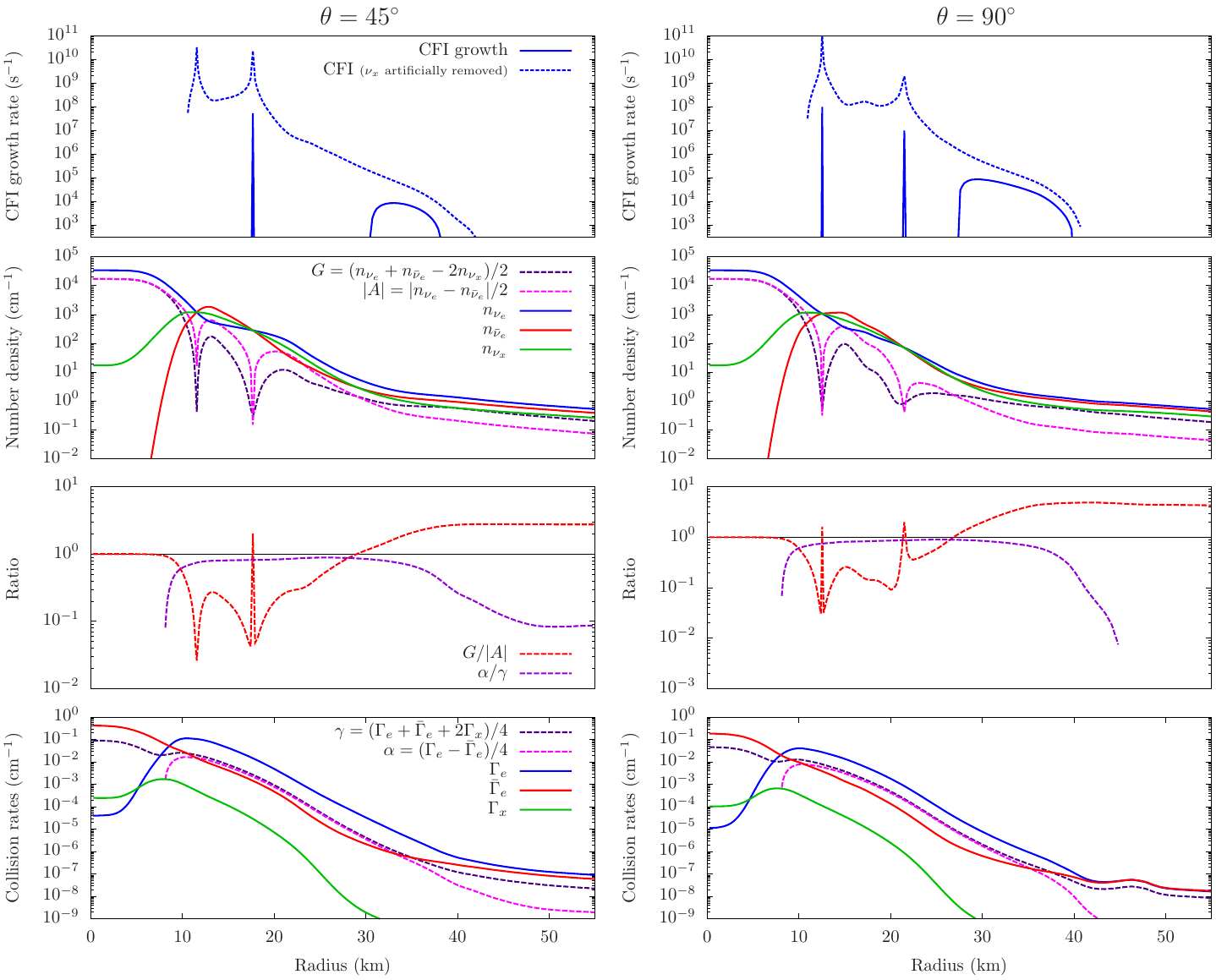}
\caption{From top to bottom, radial profiles of (1) the growth rates of CFI and FFI, (2) ratios $G/A$ and $\alpha/\gamma$, (3) number densities and $G$, $|A|$, and (4) collision rates for each species of neutrinos and $\gamma$, $\alpha$. Left and right panels are for the angle $\theta=45^\circ$ and $\theta=90^\circ$, respectively. The snapshot is at $t=404\,\mathrm{ms}$ after bounce.}
\label{fig_CFI_rad}
\end{figure*}

The CFI growth rates are shown as solid lines in the top panels of Fig.~\ref{fig_CFI_rad}, for $\theta=45^\circ$ and $90^\circ$ at $t=404\,\mathrm{ms}$. 
Both the resonance-like CFI (sharp peaks) and the non-resonance CFI ($30\lesssim r\lesssim 40\,\mathrm{km}$) are observed in both plots.
The maximum growth rate of 
$\sim10^{-3}\,\mathrm{cm}^{-1}$ is reached by the resonance-like CFI whereas the non-resonance CFI has a typical growth rate of $\sim10^{-6}\,\mathrm{cm}^{-1}$.

In the same plots we present the CFI growth rate when we artificially set the number density $\nu_x$ to zero.
In this case the CFI region is much extended, with the non-resonance CFI region merged with the resonance-like CFI region. Moreover, the growth rate becomes higher by orders with the maximum growth rate reaching $\sim1\,\mathrm{cm}^{-1}$ for the resonance-like CFI. This experiment clearly demonstrates that the existence of $\nu_x$ suppress CFI.
This is in sharp contrast with FFI, on which $\nu_x$ has no effect as long as $\nu_x$ and $\bar{\nu}_x$ do not have angular crossing. 

In the following we look into the resonance-like CFI and non-resonance CFI's more closely in turn.

\subsubsection{Resonance-like CFI}
\label{sec_resonancelike}
The resonance-like CFI occurs when the situation $A\approx0$ is realized \cite{Liu2023a,Liu2023b}.
This is vindicated in the panels in the second row of Fig. \ref{fig_CFI_rad}, where we plot the radial profiles of the number densities of all neutrinos as well as $G=(n_{\nu_e}+n_{\bar{\nu}_e}-2n_{\nu_x})/2$ and $|A|=|n_{\nu_e}-n_{\bar{\nu}_e}|/2$ (see Eq. \ref{eq_G_A}). The very sharp dips in $A$ correspond to the peaks in the growth rate (see the top panels) indeed. It is also found that $G$ has dips at the same positions, but so as deep as $A$. By definition, the situation $A\approx0$ occurs when the number densities of $\nu_e$ and $\bar\nu_e$ become close to each other. 
On the other hand, $G$ becomes zero if $n_{\nu_e} + n_{\bar{\nu}_e} = 2 n_{\nu_x}$, which is not completely the case at $A=0$. As a result, $G/|A|$ gets very large at the points, creating the resonance-like CFI as observed in the plots on the third row of Fig. \ref{fig_CFI_rad}. Note that we assume $n_{\nu_x}=n_{\bar{\nu}_x}$. If this assumption is not valid due to muonization, it may prevent $A$ to become zero at the point where $n_{\nu_e}=n_{\bar{\nu}_e}$, and might hinder the resonance-like CFI. We will investigate it in the future.

Here we comment on the possible artifact of the low radial resolution.
With a finite number of grid points, it is impossible to have $A=0$ on one of the grid points. As a result, the CFI growth rate is underestimated in the vicinity of the resonance-like CFI. The insufficient resolution also explains the absence of the resonance-like CFI at $r\sim10\,\mathrm{km}$ for $\theta=45^\circ$ in spite of $n_{\nu_e}\sim n_{\bar{\nu}_e}$. As a matter of fact, matter is more compressed and the scaleheight at this angle is shorter than at $\theta=90^{\circ}$. 

The non-detection of the resonance-like CFI in the 1D study \cite{Liu2023b} is not an artifact by the low resolution, on the other hand. As already mentioned, the abundance of $\bar{\nu}_e$ tends to be underestimated in 1D due to the lack of convection. As a result, $A=0$, which is equivalent to resonance-like CFI, is unlikely to be realized.
This clearly indicates the importance of multi-dimensionality for CFI.

\subsubsection{Non-resonance CFI}
\label{sec_nonresonance}
We now move on to the non-resonance CFI. The inner edge of the CFI region ($r\sim30\,\mathrm{km}$) corresponds to the position where $n_{\bar{\nu}_e}$ exceeds $n_{\nu_x}$. Then $G>|A|$ holds above this radius. Since $\gamma\approx\alpha$ is satisfied, it leads to the occurrence of the ordinary non-resonance CFI there.
At larger radii ($r\gtrsim40\,\mathrm{km}$), however, the CFI ceases to exist despite $G>|A|$ is sustained. This is because the ratio $\alpha/\gamma$ gets smaller as shown in the panels on the third row of Fig. \ref{fig_CFI_rad}. The two ratios $G/A$ and $\gamma/\alpha$ dictate the emergence/extinction of the CFI region: the growth rate becomes positive (and hence the CFI occurs) only when they are comparable or larger than unity. 

The behavior of $\gamma$ and $\alpha$ can be understood from the panels in the fourth row of Fig. \ref{fig_CFI_rad}, where the collision rates (Eq. \ref{eq_gamma_i}) are plotted together with $\alpha$ and $\gamma$. It is found that $\Gamma_e$ is dominant over $\bar{\Gamma}_e$ and $\Gamma_x$ at $10\lesssim r\lesssim 30\,\mathrm{km}$, which results in $\gamma\sim\alpha$ there. 
At larger radii $r\gtrsim40\,\mathrm{km}$, on the other hand, $\bar{\Gamma}_e$ becomes comparable to $\Gamma_e$. As a result, $\alpha$ gets smaller than $\gamma$.

\begin{figure*}[t] 
\includegraphics[width=\linewidth]{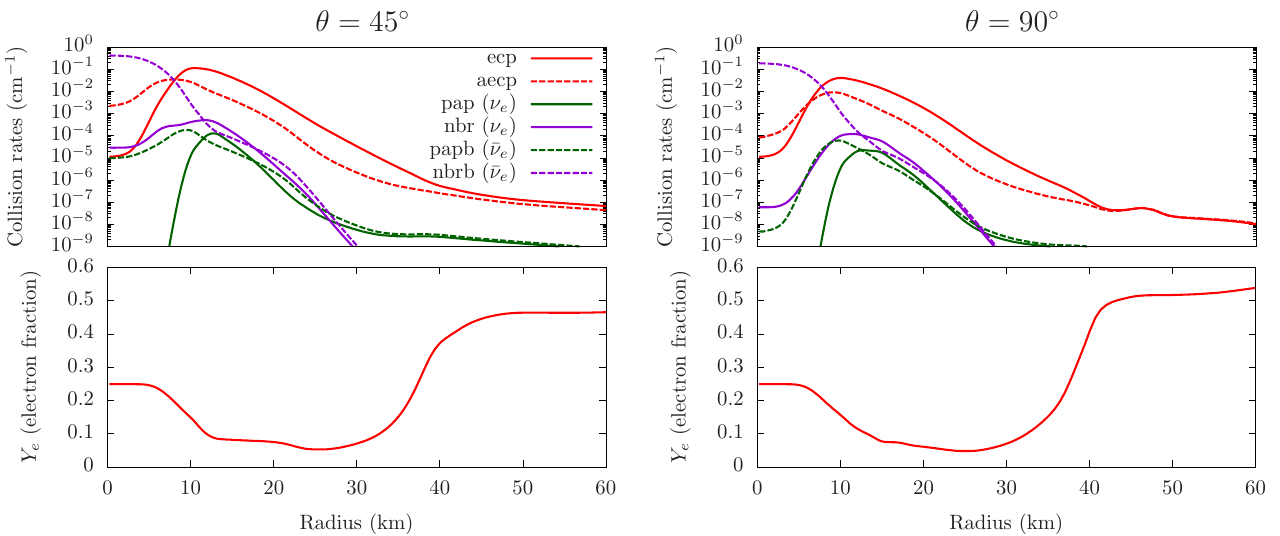}
\caption{Radial profiles of collision rates for individual neutrino interactions (top) and $Y_e$ (bottom). The angles and the time snapshot is same as Fig. \ref{fig_CFI_rad}. The abbreviations of neutrino interactions are as follows: electron capture (ecp), anti-electron capture (aecp), neutrino pair production (pap) and the nucleon bremsstrahlung (nbr).}
\label{fig_indiv}
\end{figure*}
In order to understand the behavior of $\Gamma_e$ and $\bar{\Gamma}_e$ further, we plot the contributions of individual neutrino-matter interactions in Fig. \ref{fig_indiv}. As can be seen, the electron capture on proton (ecp) and the anti-electron capture on neutron (aecp) dominate other interactions at $r \gtrsim 10\,\mathrm{km}$, which means that they mainly drive $\Gamma_e$ and $\bar\Gamma_e$, respectively. 

It is interesting to compare $\Gamma$'s with the $Y_e$ distribution shown in the bottom panels of Fig. \ref{fig_indiv}.
At $10\lesssim r \lesssim30\,\mathrm{km}$, $Y_e$ is low $\sim0.1$. 
This corresponds to the region where $\nu_e$ opacity dominates over $\bar{\nu}_e$, i.e., $\Gamma_e > \bar{\Gamma}_e$.
On the other hand, at $r\gtrsim40\,\mathrm{km}$, $Y_e$ is $\sim0.5$. In this region, ecp and aecp have similar collision rates, which yields $\Gamma_e\sim\bar{\Gamma}_e$.
This analyses is in line with the 1D result that CFI was observed only for rather low-$Y_e$ region.

\subsection{FFI}
\label{sec_FFI}
\begin{figure*}[t] 
\includegraphics[width=\linewidth]{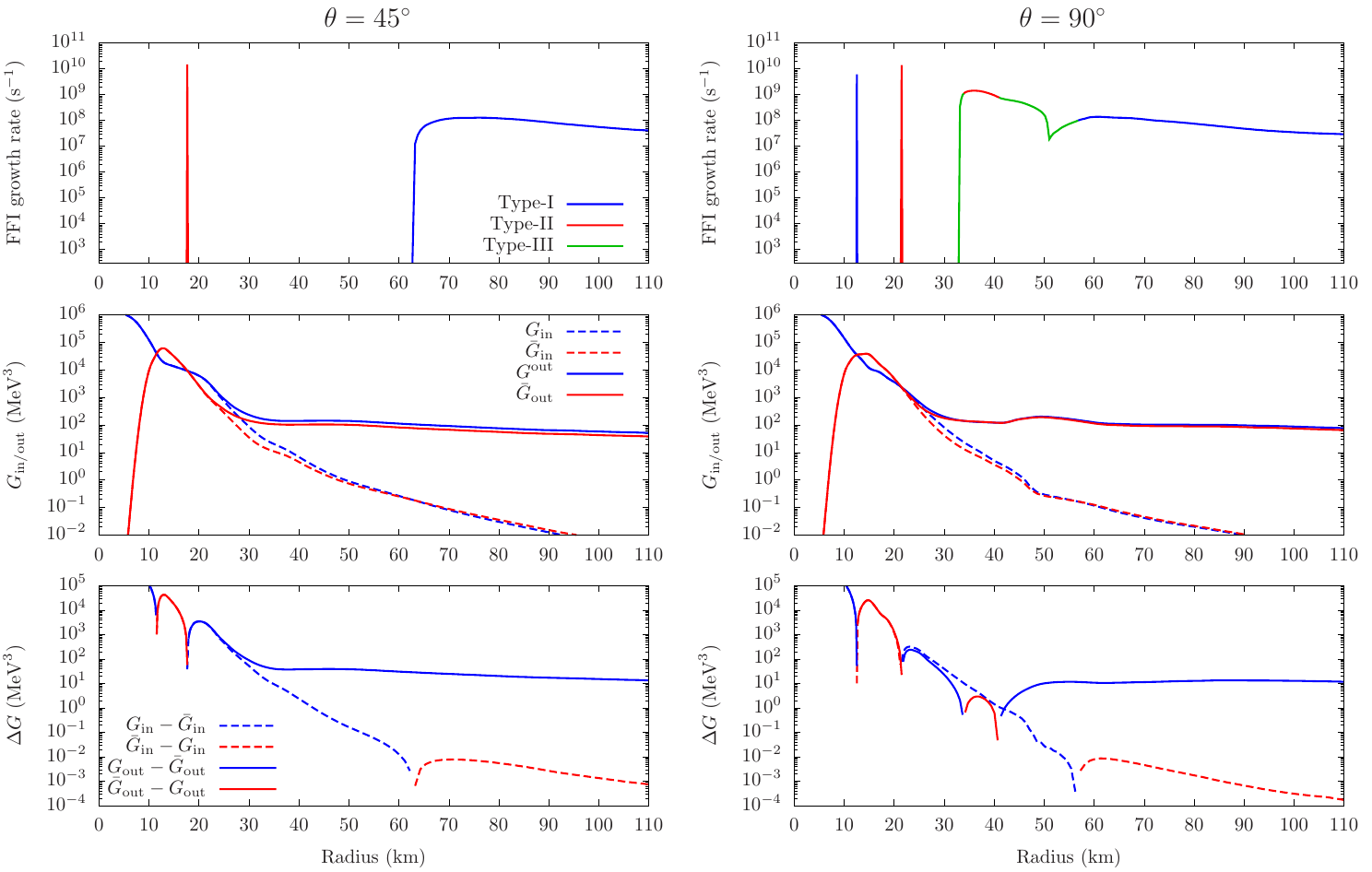}
\caption{Radial profiles of the FFI growth rate (top) and $G_{\mathrm{in/out}}$, $\bar{G}_{\mathrm{in/out}}$ (middle) and the differences between $G_{\mathrm{in/out}}$ and $\bar{G}_{\mathrm{in/out}}$ (bottom) at $t=404\,\mathrm{ms}$ after bounce.}
\label{fig_FFI_rad}
\end{figure*}
We turn our attention to the FFI region found in our model. 
Top panels of Fig. \ref{fig_FFI_rad} shows the growth rates of FFI for $\theta=45$ and $90^\circ$.
Different colors distinguish the types of angular crossing. Here, we use the terminology of \cite{Nagakura2021b}; type-I crossing means $\nu_e$ is dominant over $\bar{\nu}_e$ in the outgoing direction ($\mu_\nu=1$) whereas $\bar{\nu}_e$ is dominant over $\nu_e$ in the incoming direction ($\mu_\nu=-1$). Type-II crossing means the opposite. Note that it is possible that FFI exists but the type cannot be categorized into either of them. We call this case type-III hereafter. There are two possible reasons; (1) the number of crossing is even, or (2) shallow crossing appears for some energy or $\phi_\nu$, but the integration smear it out. Note that we judge the crossing type by the energy-integrated and $\phi_\nu$-averaged distribution function. 
It should be pointed out that the detection scheme proposed previously for results obtained with the truncated moment method \cite{Nagakura2021b} assumed an odd number of crossings. The FFI region with even number of crossings may have been overlooked with such a scheme.

\begin{figure}[t] 
\includegraphics[width=\linewidth]{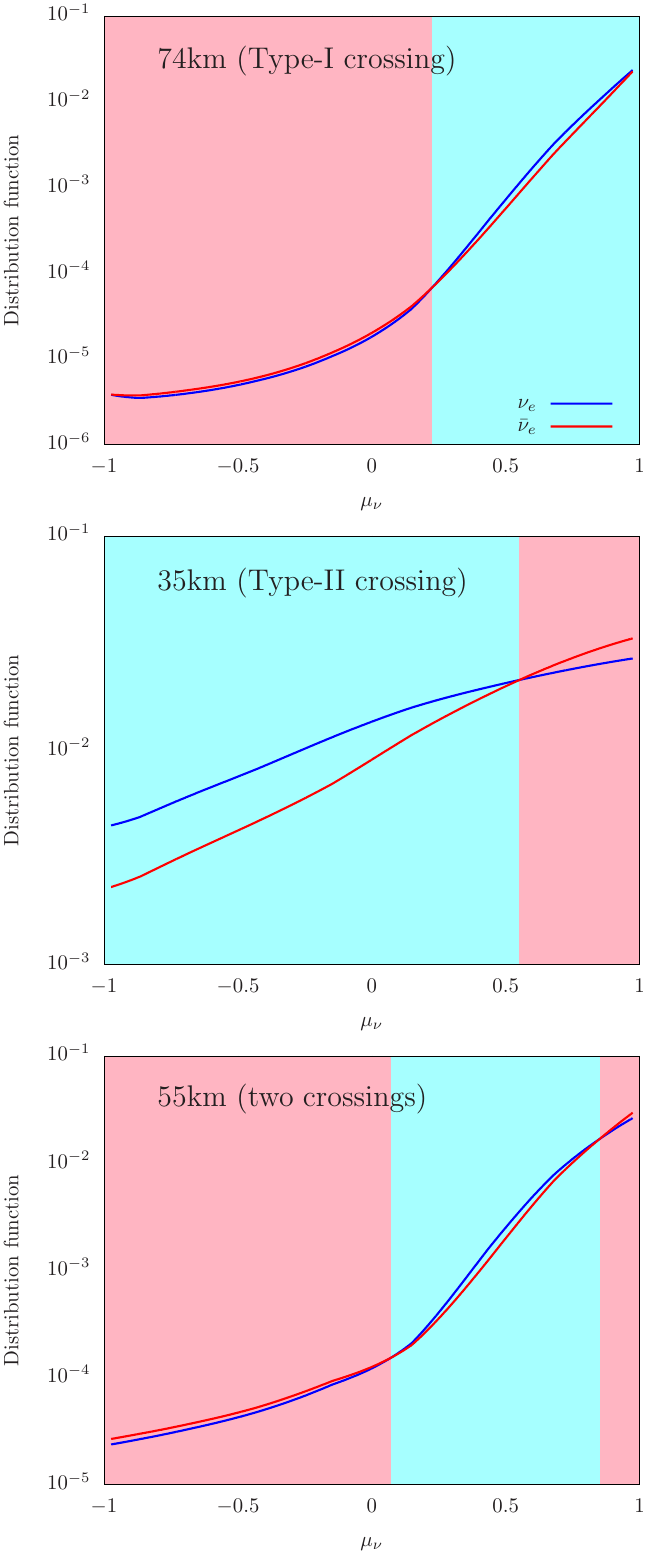}
\caption{Angular distribution of the distribution function of $16.5\mathrm{MeV}$ neutrinos at $74\,\mathrm{km}$ (top), $35\,\mathrm{km}$ (middle), $55\,\mathrm{km}$ (bottom). The angle is $\theta=90^\circ$ and the snapshot time is $t=404\,\mathrm{ms}$ after bounce.}
\label{fig_angdis}
\end{figure}

In the middle panels of Fig.~\ref{fig_FFI_rad}, we exhibit the energy-integrated and $\phi_\nu$-averaged distribution functions defined as
\begin{eqnarray}
&&G_{\mathrm{in}}\equiv\int E^2dE\int\frac{d\phi_\nu}{2\pi}f_{\nu_e}(\mu_\nu=-1)\quad (\mathrm{incoming}), \\
&&G_{\mathrm{out}}\equiv\int E^2dE\int\frac{d\phi_\nu}{2\pi}f_{\nu_e}(\mu_\nu=1)\quad (\mathrm{outgoing}),
\end{eqnarray}
and $\bar{G}_{\mathrm{in/out}}$ for $\bar{\nu}_e$. The bottom panels give the difference of ${G}_{\mathrm{in/out}}$ between $\nu_e$ and $\bar{\nu}_e$. 
Since it is a logarithmic plot, a line is shown only if the value is positive.
Note that the colors distinguish the signatures. 
The combination of blue solid line and red dashed line means type-I crossing exists there, whereas the pair of blue dashed line and red solid line corresponds to type-II crossing. Other combinations indicate either no crossing or type-III crossing.
There are several types and reasons of FFI in the semi-transparent and optically-thin region. The angular distributions for three representative radii are shown in Fig. \ref{fig_angdis}.

Type-I crossing is observed at $r\gtrsim60\,\mathrm{km}$ for both angles (see Fig. \ref{fig_FFI_rad} and the top panel of Fig. \ref{fig_angdis}). As already pointed out in \cite{Nagakura2021b}, this is produced by the back-scattering of $\bar{\nu}_e$. Since $\bar{\nu}_e$'s tend to have higher energies than $\nu_{e}$'s as they come from deeper inside, the nucleon scattering occurs more frequently for $\bar{\nu}_e$ than for ${\nu}_e$. It produces a larger population of the former in the inward direction. It is mentioned that type-I crossing produced this way was observed only for the exploding models in \cite{Nagakura2021b}.

We find a type-II crossing at $r\sim40\,\mathrm{km}$ only for $\theta=90^\circ$ (also see the middle panel of Fig. \ref{fig_angdis}). It actually corresponds to the mushroom-shaped FFI region in Fig. \ref{fig_2Dslices}, which is produced by convective motions.
As may be inferred from the $G_{\mathrm{in/out}}$ distributions in Fig. \ref{fig_FFI_rad} , it is located in the neutrino decoupling region.
Because $\bar{\nu}_e$ decouples from matter deeper inside than ${\nu}_e$, its angular distribution in momentum space is more forward-peaked. As a result, $\bar{\nu}_e$ is more abundant than ${\nu}_e$ for the outgoing direction while the opposite is true for the incoming direction. The generation of type-II crossing by this mechanism was already discussed in previous studies \cite{Nagakura2021b,Harada2022,Akaho2023}.
It did not happen in other angles including $\theta=45^\circ$, because ${\nu}_e$ is clearly dominant over $\bar{\nu}_e$ there. 

Type-III crossings are found at $\theta=90^\circ$. They are actually separated into two regions; (1) the very narrow strip at the inner boundary of type-II crossing, and (2) the domain between type-I and type-II crossing regions. The former corresponds to the shallow crossing mentioned earlier.
On the other hand, the latter domain has two crossings instead of one. The typical angular distribution is presented in the bottom panel of Fig. \ref{fig_angdis}. In fact, We find that $\bar{\nu}_e$ is dominant over ${\nu}_e$ at both $\mu_\nu=-1$, $1$ but opposite for $\mu_\nu\sim0$.
Since this domain is sandwiched by the type-II crossing region at smaller radii and the type-I crossing region at larger radii, both mechanisms operate in this region, creating the two crossings.
As mentioned earlier, the detection of FFI based on moments assuming that the number of crossings is odd \cite{Nagakura2021b} will fail to find this region. In this respect, the Boltzmann neutrino transfer is certainly advantageous.

We find very narrow spikes in the FFI growth rate at both $\theta = 45^\circ$ and $90^\circ$. They are located at the same position as the resonance-like CFI, as we will see later. This is natural because the condition $n_{\bar{\nu}_e}/n_{\nu_e}\sim1$ is favorable not only for the resonance-like CFI but also for the FFI, as already reported previously \cite{Glas2020,DelfanAzari2020}.
The absence of the inner peak for $\theta=45^\circ$ is due to the low radial resolution, just as for the resonance-like CFI.
The type of crossing at this point is rather meaningless because both $\nu_e$ and $\bar{\nu}_e$ have almost isotropic distributions at these points.

\subsection{Comparison between CFI and FFI}
\label{sec_comparison}
\begin{figure}[t] 
\includegraphics[width=\linewidth]{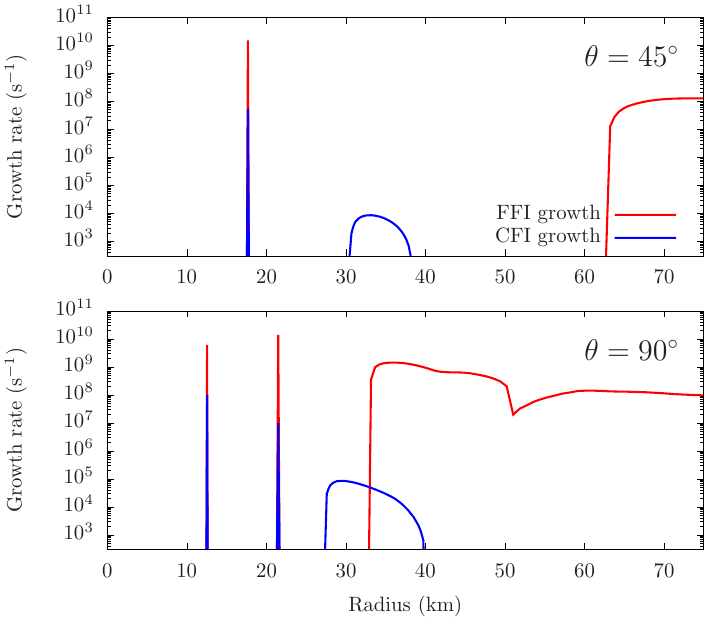}
\caption{Radial profiles of the CFI and FFI growth rates for $\theta=45^\circ$ (top) and $\theta=90^\circ$ (bottom) at $t=404\,\mathrm{ms}$ after bounce.}
\label{fig_rad_comp}
\end{figure}
Finally, we compare the growth rates of CFI and FFI in Fig. \ref{fig_rad_comp}. It is clear that the growth rate of FFI is higher than CFI by many orders if both of them exist. 
This is as expected because the dependence of the growth rate on the neutrino number density $n$ is different between the two modes; $\sigma_\mathrm{FFI}\propto n$, and $\sigma_\mathrm{CFI}\propto \sqrt{n}$. 
However, it is worth mentioning that the relation $\sigma_\mathrm{FFI}\gg\sigma_\mathrm{CFI}$ is not the universal relationship and may be opposite if the angular crossing is shallow or the collision rates are large.

The above comparison indicates that CFI is subdominant in the linear evolution even the resonance-like CFI occurs. 
However, it does not mean that CFI is unimportant. As long as the growth rate is shorter than the typical time scale of the background evolution, the flavor conversion will reach the nonlinear phase anyway. The subsequent evolution and possible saturation are currently under extensive investigations \cite{Padilla-Gay2022,Johns2022,Lin2023,Xiong2023a,Xiong2023c}. For example, the Monte Carlo simulations in \cite{Kato2023b} found that the resonance-like CFI induces the flavor swap rather than the settlement to the flavor equilibrium. It will be eventually needed to somehow incorporate these results in the supernova simulations and see their effects on the fluid dynamics, neutrino signals, and nucleosynthesis in CCSNe.

\section{Conclusions.}
\label{sec_conclusion}
We conducted the post-process analyses of one of our 2D CCSN
simulations performed with the Boltzmann neutrino radiation hydrodynamics code to search for the regions where the collisional and/or fast flavor instabilities will possibly happen. We employed the criterion for these flavor instabilities that were derived in the previous studies \cite{Morinaga2020,Liu2023a} based on the linear analysis.

We found that the non-resonance CFI would occur in the region with the density of $10^{10}\lesssim \rho\lesssim 10^{12}\,\mathrm{g}\,\mathrm{cm}^{-3}$, which is consistent with the previous findings in the 1D study \cite{Liu2023b}. In the multi-dimensional model, however, the radial extent of the CFI region changes in time on the dynamical timescale, which was absent in the 1D model. This is due to the turbulence in the supernova core. 
Non-resonance CFI region is likely to be separated from FFI region most of the time, but they can be overlapped with each other
at some angles depending on the asymmetry of fluid motions.
The non-resonance CFI region is characterized as follows;
the inner boundary corresponds to the points where the number density of $\bar{\nu}_e$ becomes equal to that of $\nu_x$ i.e., $G = |A|$. On the other hand, the outer boundary corresponds to the positions where 
$\bar{\nu}_e$ opacity becomes comparable to that of $\nu_e$.
It is also noted that the outer edge roughly corresponds to $Y_e\approx0.5$.

We found that the resonance-like CFI occurs when the value of $A$ is close to zero, which happens in turn if the number densities of different species of neutrinos almost coincide with one another. This is in contrast with the previous 1D study \cite{Liu2023b}. As mentioned earlier, abundance of $\bar{\nu}_e$ tend to be artificially suppressed in 1D, which makes it hard to realize $A\approx0$. Our result clearly indicate the importance of multi-dimensional effect for CFI.

The overall properties of the appearance of FFI regions we observed in this study are consistent with those of the previous study in \cite{Nagakura2021b}; (1) in the optically thick region, the FFI occurs if $n_{\nu_e}/n_{\bar{\nu}_e}\sim1$ , (2) in the decoupling region, type-II crossing occurs if $\bar{\nu}_e$ emission is strong, and (3) in the optically thin region, type-I crossing is produced due to nucleon scattering. 
However, we found that multiple angular crossing can be realized in the domain between the regions with type-I and type-II crossings. Note that this detection was made possible by the exploitation of the results of Boltzmann neutrino transport, where the full information on the angular distribution in momentum space is available. 

The linear growth rate of CFI is always lower than that of FFI by many orders. This is true of the resonance-like CFI also but its growth rate is larger than that of the non-resonance counterpart by orders. It should be pointed out that whether CFI or FFI have larger linear growth rates may not be so important. As a matter of fact, as long as they are shorter than the typical timescale of background evlolutions and the neutrino crossing time over the background scaleheight, the flavor conversions reach the nonlinear stage anyway. The eventual outcomes should then be explored with different approaches \cite{Johns2022,Padilla-Gay2022,Zaizen2022,Hansen2022,Kato2023b}.

We wrap up this paper by noting the limitations of this study and giving some future prospects. 
First, as we have just mentioned, this study is based on the linear analysis, which can address only the trigger of flavor conversions. The subsequent evolution and the asymptotic state should be investigated, for example, by directly solving the QKE. 
\highlight{Second, flavor conversions at a certain spatial position propagate in space, leading to a qualitative change of global neutrino radiation field in CCSNe
(\cite{Nagakura2022,Shalgar2023,Nagakura2023a,Nagakura2023d}).
However, our post-process analysis does not have the ability to incorporate the feedback of global neutrino advection, which should be kept in mind as a caveat.
We are updating our Boltzmann radiation-hydrodynamics code to incorporate the possible outcomes of FFI and CFI and the results will be reported in the future.}

\clearpage
\begin{acknowledgments}
This research used the K and Fugaku supercomputers provided by RIKEN, the FX10 provided by Tokyo University, the FX100 provided by Nagoya University, the Grand Chariot provided by Hokkaido University, and Oakforest-PACS provided by JCAHPC through the HPCI System Research Project (Project ID: hp130025, 140211, 150225, 150262, 160071, 160211, 170031, 170230, 170304, 180111, 180179, 180239, 190100, 190160, 200102, 200124, 210050, 210051, 210164, 220047, 220173, 220047, 220223 and 230033), and the Cray XC50 at Center for Computational Astrophysics, National Astronomical Observatory of Japan (NAOJ). This work is supported by 
Grant-in-Aid for Scientific Research
(19K03837, 20H01905,21H01083)
and 
Grant-in-Aid for Scientific Research on Innovative areas 
"Gravitational wave physics and astronomy:Genesis" 
(17H06357, 17H06365) and ”Unraveling the History of the Universe and Matter Evolution with Underground Physics” (19H05802 and 19H05811)
from the Ministry of Education, Culture, Sports, Science and Technology (MEXT), Japan.
For providing high performance computing resources, Computing Research Center, KEK, JLDG on SINET of NII, Research Center for Nuclear Physics, Osaka University, Yukawa Institute of Theoretical Physics, Kyoto University, Nagoya University, and Information Technology Center, University of Tokyo are acknowledged. This work was supported by 
MEXT as "Program for Promoting Researches on the Supercomputer Fugaku" 
(Toward a unified view of the universe: from large scale structures to planets, JPMXP1020200109) and the Particle, Nuclear and Astro Physics Simulation Program (Nos. 2020-004, 2021-004, 2022-003) of Institute of Particle and Nuclear Studies, High Energy Accelerator Research Organization (KEK).
RA is supported by JSPS Grant-in-Aid for JSPS Fellows (Grant No. 22J10298) from MEXT.
HN is supported by Grant-in Aid for Scientific Research (23K03468).
MZ is supported by JSPS Grant-in-Aid for JSPS Fellows (No. 22KJ2906) from MEXT.
SY is supported by Institute for Advanced Theoretical and Experimental Physics, Waseda University, and the Waseda University Grant for Special Research Projects (project No. 023C-141).
\end{acknowledgments}

%\nocite{*}

\bibliography{apssamp}% Produces the bibliography via BibTeX.

\end{document}